\documentclass{article}
\usepackage{spconf,amsmath,graphicx}
\graphicspath{ {./figs/} }
\usepackage{booktabs}
\usepackage{enumitem}
\setlist{nosep, leftmargin=14pt}

\usepackage{mwe} 

\title{TABSurfer: a Hybrid Deep Learning Architecture for Subcortical Segmentation}
\name{Aaron Cao$^{1}$, Vishwanatha M. Rao$^{2}$, Kejia Liu$^{2}$, Xinrui Liu$^{3}$, Andrew F. Laine$^{2}$, Jia Guo$^{4,5,*}$ \thanks{*Correspondence: Jia Guo, jg3400@columbia.edu}}

\address{$^{1}$ Valley Christian High School, San Jose, CA, USA \\
    $^{2}$ Department of Biomedical Engineering, Columbia University, New York, NY, USA \\
    $^{3}$ The Village School, Houston, TX, USA \\
    $^{4}$ Department of Psychiatry, Columbia University, New York, NY, USA \\
    $^{5}$ Mortimer B. Zuckerman Mind Brain Behavior Institute, Columbia University, New York, NY, USA}

\begin{document}
%
\maketitle
\begin{abstract}

Subcortical segmentation remains challenging despite its important applications in quantitative structural analysis of brain MRI scans. The most accurate method, manual segmentation, is highly labor intensive, so automated tools like FreeSurfer have been adopted to handle this task. However, these traditional pipelines are slow and inefficient for processing large datasets. In this study, we propose TABSurfer, a novel 3D patch-based CNN-Transformer hybrid deep learning model designed for superior subcortical segmentation compared to existing state-of-the-art tools. To evaluate, we first demonstrate TABSurfer's consistent performance across various T1w MRI datasets with significantly shorter processing times compared to FreeSurfer. Then, we validate against manual segmentations, where TABSurfer outperforms FreeSurfer based on the manual ground truth. In each test, we also establish TABSurfer's advantage over a leading deep learning benchmark, FastSurferVINN. Together, these studies highlight TABSurfer's utility as a powerful tool for fully automated subcortical segmentation with high fidelity.

\end{abstract}
\begin{keywords}
Biomedical Image Processing, Deep Learning, Semantic Segmentation
\end{keywords}
\section{Introduction}
\label{sec:intro}

Subcortical segmentation is a significant application in medical image processing, extracting quantitative structural information on subcortical regions within an MRI scan. This can aid in detecting and tracking morphological deficits in various neuropsychiatric conditions, including Major Depressive Disorder \cite{ho2022subcortical}, Dementia \cite{van2023subcortical}, and Schizophrenia \cite{gutman2022meta}. 

While manual segmentation stands as the most trusted method, it is a laborious and difficult task, even for experts. Thus, computer tools like FreeSurfer \cite{fischl2002whole} have been developed to automate the process. But while FreeSurfer is now a widely accepted standard, it is inconvenient for processing large and diverse datasets. FreeSurfer's automatic subcortical segmentation can take many hours to complete for a single scan, and its traditional approach can be sensitive to data quality issues. 

Artificial intelligence and supervised deep learning approaches have recently emerged as both fast and accurate tools for semantic segmentation tasks. In particular, Convolutional Neural Network (CNN) architectures like the UNet \cite{ronneberger2015u} \cite{cciccek20163d} have become a dominant choice for medical image segmentation. With the use of GPUs, these tasks can now take just a few seconds or minutes to complete, instead of hours. However, subcortical segmentation has remained a difficult task due to the complex 3D structures within the brain, the large number of labels, and the expensive hardware memory requirements for processing scans at full resolution.

One of the leading deep learning-based alternatives to FreeSurfer is the FastSurfer pipeline \cite{henschel2020fastsurfer}, which includes whole brain segmentation. As a benchmark for our study, we evaluate our model against their pretrained FastSurferVINN model \cite{henschel2022fastsurfervinn}, which aggregates three 2D F-CNNs for a 2.5D approach. However, the 2D models within FastSurferVINN inevitably struggle to fully capture the complex 3D spatial dependencies within the anatomical structures of the brain. 

On the other hand, 3D patch-based solutions are better suited to capture such geometries. While full 3D volume deep learning models for segmenting many classes are currently not possible due to data and memory constraints, a patch-based approach is less computationally expensive, while also generating more training samples per subject and better capturing local 3D information. However, utilizing these patches sacrifices global context by focusing on a local view.

Recently, Transformers have demonstrated state-of-the-art performance in natural image segmentation. While CNN variations have outperformed previous machine learning algorithms in this task, evidence has emerged of further improved generalization and performance by coupling Transformers with CNNs \cite{chen2021transunet} \cite{hatamizadeh2022unetr}. 

With these insights, we propose TABSurfer, a new deep learning model inspired by the TABS architecture \cite{rao2022improving}, which previously demonstrated strong performance in brain tissue segmentation. Improving on TABS's volume-based approach, we adapt the concept into a 3D patch-based implementation, focusing on the task of subcortical segmentation for 31 regions (all of the subcortical structures covered by FastSurferVINN excluding left and right cortical white matter). The model roughly resembles a ResUnet \cite{zhang2018road}, but with a Vision Transformer module as the bridge connecting the encoder and decoder paths to extract more context and compensate for the limitations of working on local patches.

In this study, we evaluate the performance of TABSurfer compared to both the well-established FastSurfer and FreeSurfer segmentation tools, showing the effectiveness of new hybrid architectures and Transformers for handling complex segmentations containing many classes.

\section{Materials and Methods}
\label{sec:format}

\subsection{Data}
\label{ssec:subhead}
\begin{figure}[htb]
\centering
  \begin{minipage}[htb]{0.66\linewidth}  
    \centering
    \includegraphics[width=6.25cm]{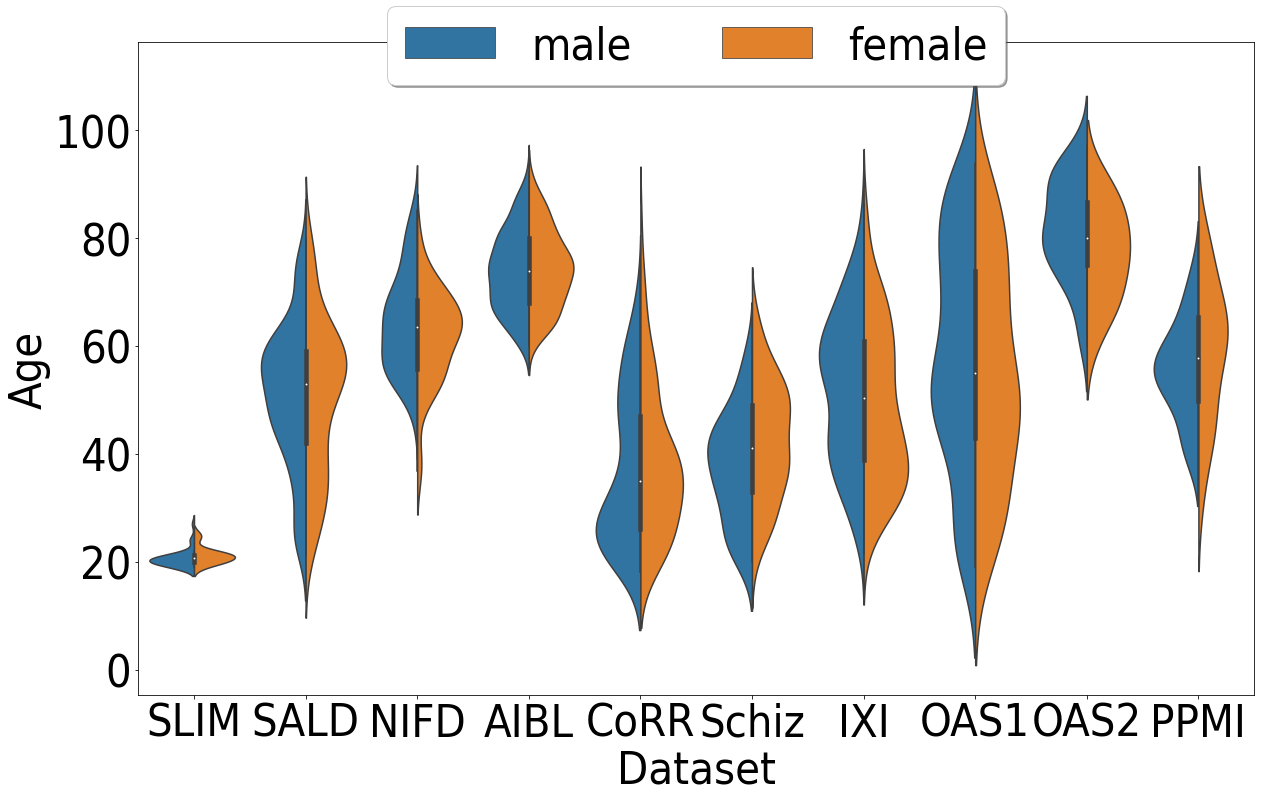}
  \end{minipage}
  \hfill
  \begin{minipage}[htb]{0.33\linewidth}  
    \centering
    \includegraphics[width=1.8cm]{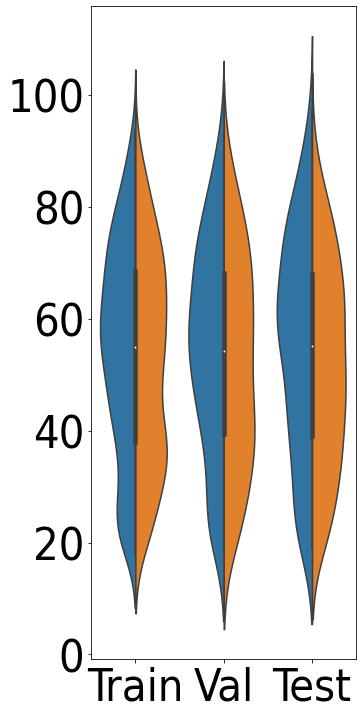}
  \end{minipage}
  \caption{Age and Gender Distributions for each Dataset}
  \label{fig:data}
\end{figure}

We selected 1788 T1w MRI scans from a large-scale heterogeneous dataset assembled from various publicly available sources \cite{feng2020estimating}. This data was divided into training, validation, and test sets with a roughly 3:1:1 ratio. The training set had 1079 scans, the validation set had 345, and the test set had 364. We achieve a balanced age and gender distribution between the diverse selection of datasets, as shown in Figure~\ref{fig:data}. Ground truth segmentations for this data were generated using FreeSurfer. Additionally, we obtained 20 manually segmented scans from the MindBoggle-101 OASIS-TRT-20 dataset \cite{klein2012101}. Five of these were added to the training set and the rest were used as ground truths for a separate test set. The T1w scans were preprocessed with skull-stripping and intensity normalization to create the inputs for our models.

\subsection{Pipeline and Model Architecture}
\label{ssec:subhead}
\begin{figure*}
  \centering
  \includegraphics[width=1\textwidth]{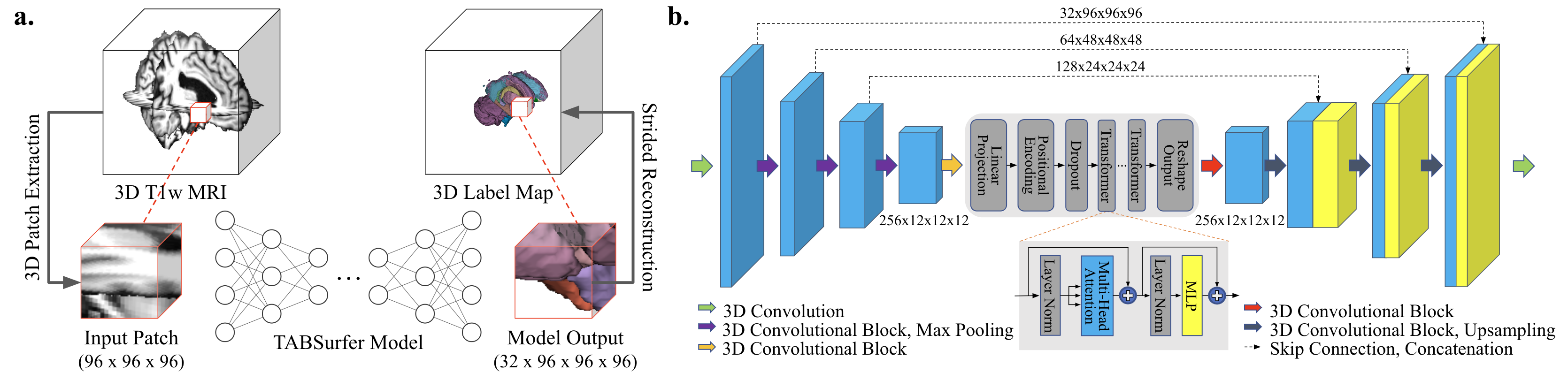}
  \caption{\textbf{a.} Our pipeline extracts 3D patches from the input scan, feeds them into our model, and reconstructs the output predicted classes to generate a segmentation. \textbf{b.} Visualization of the TABSurfer model's architecture.}
  \label{fig:pipeline+architecture}
\end{figure*}

Our pipeline follows a 3D patch-based approach with a hybrid CNN-Transformer model, as visualized in Figure~\ref{fig:pipeline+architecture}. 

First, the input scan is centered and conformed to RAS orientation, and the intensities are rescaled from 0 to 1 in the same way as in FastSurfer's pipeline. This input volume with dimensions 256 x 256 x 256 is cropped and padded before patch extraction. Each patch has dimensions 96 x 96 x 96, and we set the step size between each patch to 16. Each patch is fed into the model sequentially, and the output class probabilities are reconstructed to the shape of the original input image. Each patch's predicted probabilities are combined to vote on the class for each voxel, and the values are then mapped to the corresponding FreeSurfer label. This pipeline ensures that the model can segment an entire scan in less than 90 seconds.

Our model architecture consists of a 3D CNN encoder and decoder with skip connections, and a Vision Transformer module in between. Passing through the encoder, four layers of residual blocks and max pooling operations downsample the input patch for an encoded feature tensor. Using “linear projection and learned positional embedding” operations \cite{wang2021transbts}, we convert the encoded feature tensor into 1024 tokenized vectors. These are sequentially fed into the Transformer encoder \cite{vaswani2017attention}, which consists of 8 layers and 16 heads. The reshaped output of the Transformer is then passed to the decoder, which reconstructs the image to the original input dimensions. Finally, a convolution operation and a Softmax activation function are applied to generate a 32-channel output, where each channel corresponds to the probability for an individual class. Each residual block within the encoder and decoder layers consists of a residual connection and a sequence of 3D Convolution, Group Normalization, and Rectified Linear Unit (ReLU).

\subsection{Model Training}
\label{ssec:subhead}
The model described above was trained on a 24 GB NVIDIA Quadro 6000 GPU. We utilized the AdamW optimizer with a learning rate of 1e-6 and a weight decay of 1e-4. We applied three forms of augmentation with a probability of 0.2 each: affine, noise, and blur. Our loss function was Dice Loss. 

\subsection{Model Evaluation}
\label{ssec:subhead}
We conducted two tests to evaluate our model's performance. First, we evaluated TABSurfer against FastsurferVINN (from the FastSurfer Github) using 364 FreeSurfer segmentations as ground truths. Second, we validated TABSurfer against both FastSurferVINN and FreeSurfer on 15 manual segmentations as ground truths.

We used the Dice Similarity Coefficient (DSC) and the Average Symmetric Surface Distance (ASSD) metrics to evaluate both the overall similarity of the segmentations and the quality of the contours against the ground truth.

\section{Results}
\label{sec:pagestyle}

\subsection{Evaluation on FreeSurfer Segmentations}
\label{ssec:subhead}

Average metrics from evaluating TABSurfer and FastSurferVINN against the FreeSurfer-generated ground truth are displayed in Table~\ref{table:freesurfer}. TABSurfer consistently achieved high Dice Similarity Coefficient scores, with the mean for each dataset never falling below 0.85 and reaching above 0.87 on most datasets. In contrast, the benchmark, FastSurferVINN, struggled with inconsistent performance, reaching an average Dice Similarity Coefficient as low as 0.812 on the IXI dataset.

\begin{figure*}
  \centering
  \includegraphics[width=.85\textwidth]{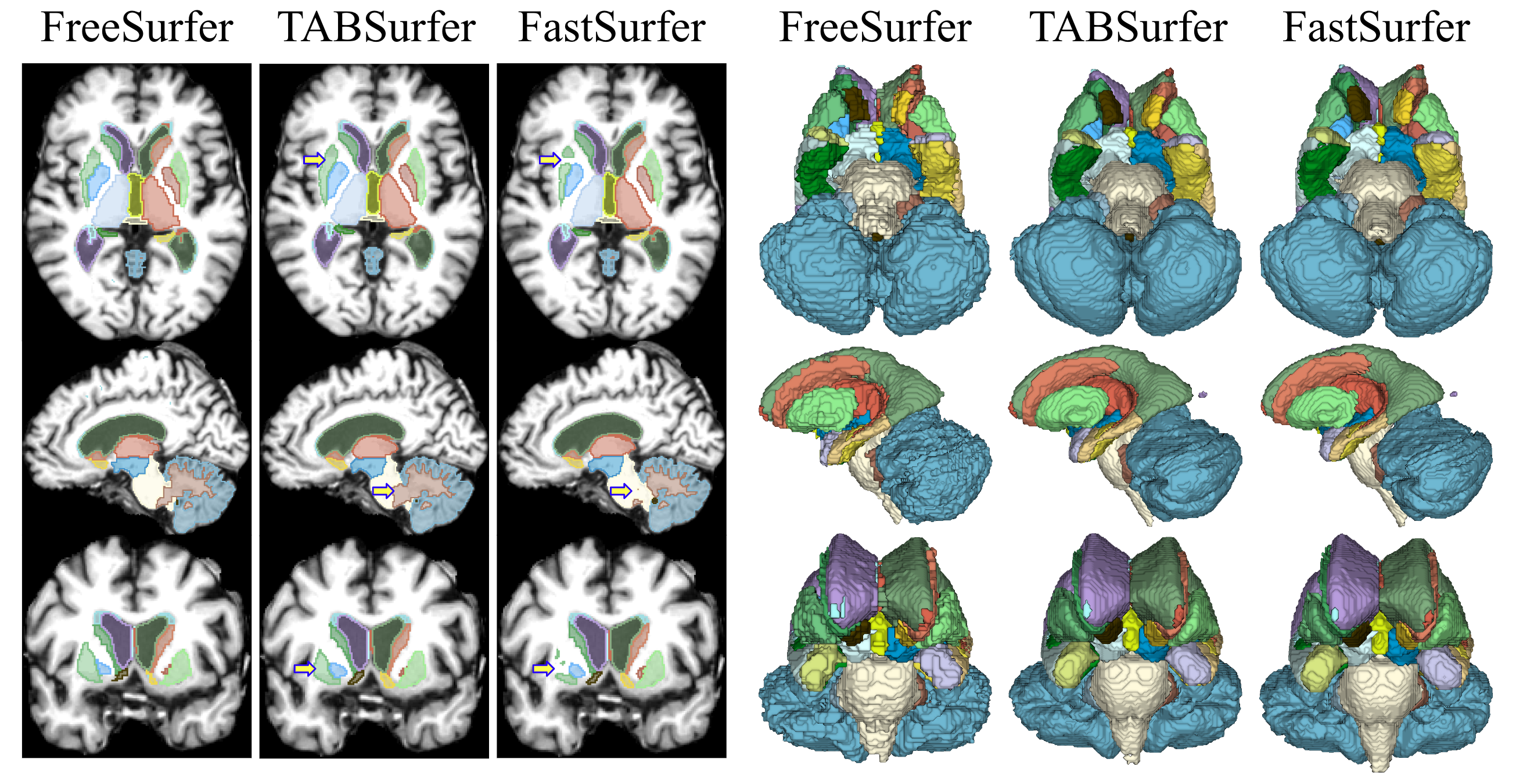}
  \caption{Sample predicted slices and volumes between FreeSurfer, TABSurfer, and FastSurferVINN segmentations}
  \label{fig:qualitative}
\end{figure*}

The visualization of sample segmentations in Figure~\ref{fig:qualitative} also reveals TABSurfer’s increased image quality over both FreeSurfer and FastSurferVINN. TABSurfer captures each structure more fully compared to FastSurferVINN, while obtaining smoother contours compared to FreeSurfer.

\begin{table}[htb]
\centering
\small
\begin{tabular}{llll}
\toprule
Dataset &      Model &          DSC ↑ &           ASSD ↓ \\
\midrule
   AIBL &  \textbf{TABSurfer} & \textbf{0.887 ± 0.010} & \textbf{0.318 ± 0.046} \\
    & FastSurfer & 0.879 ± 0.015 & 0.335 ± 0.059 \\
   \hline
   CoRR &  \textbf{TABSurfer} & \textbf{0.875 ± 0.022} & \textbf{0.358 ± 0.087} \\
    & FastSurfer & 0.866 ± 0.027 & 0.380 ± 0.104 \\
   \hline
    IXI &  \textbf{TABSurfer} & \textbf{0.853 ± 0.028} & \textbf{0.471 ± 0.125} \\
     & FastSurfer & 0.812 ± 0.034 & 0.614 ± 0.140 \\
    \hline
   NIFD &  \textbf{TABSurfer} & \textbf{0.889 ± 0.009} & \textbf{0.304 ± 0.038} \\
    & FastSurfer & 0.888 ± 0.008 & 0.305 ± 0.030 \\
   \hline
   OAS1 &  \textbf{TABSurfer} & \textbf{0.879 ± 0.012} & \textbf{0.339 ± 0.044} \\
    & FastSurfer & 0.875 ± 0.010 & 0.341 ± 0.047 \\
   \hline
   OAS2 &  \textbf{TABSurfer} & \textbf{0.880 ± 0.012} & 0.332 ± 0.046 \\
    & \textbf{FastSurfer} & 0.880 ± 0.013 & \textbf{0.324 ± 0.049} \\
   \hline
   PPMI &  \textbf{TABSurfer} & \textbf{0.886 ± 0.010} & \textbf{0.319 ± 0.039} \\
    & FastSurfer & 0.879 ± 0.008 & 0.328 ± 0.033 \\
   \hline
   SALD &  \textbf{TABSurfer} & \textbf{0.865 ± 0.027} & \textbf{0.399 ± 0.094} \\
    & FastSurfer & 0.842 ± 0.021 & 0.482 ± 0.089 \\
   \hline
  Schiz &  \textbf{TABSurfer} & \textbf{0.870 ± 0.011} & \textbf{0.374 ± 0.049} \\
   & FastSurfer & 0.838 ± 0.022 & 0.485 ± 0.094 \\
  \hline
   SLIM &  \textbf{TABSurfer} & \textbf{0.878 ± 0.006} & \textbf{0.333 ± 0.026} \\
    & FastSurfer & 0.855 ± 0.012 & 0.425 ± 0.051 \\
   \hline
   Full &  \textbf{TABSurfer} & \textbf{0.872 ± 0.023} & \textbf{0.374 ± 0.099} \\
    & FastSurfer & 0.854 ± 0.035 & 0.436 ± 0.143 \\
\bottomrule
\end{tabular}

\begin{minipage}{7.5 cm}
\vspace{.1 cm}
    \footnotesize Bold text indicates superior performance. Up arrow indicates that higher numbers correspond to better performance and down arrow indicates that lower numbers correspond to better performance.
\end{minipage}
\caption{Comparing TABSurfer and FastsurferVINN metrics across datasets.}
\label{table:freesurfer}
\end{table}



\subsection{Evaluation on Manual Segmentations}
\label{ssec:subhead}

Results from evaluating TABSurfer, FastSurferVINN, and FreeSurfer compared to the manual reference are shown in Table~\ref{table:manual}. FreeSurfer exhibited the poorest performance, and FastSurferVINN was marginally better; however, TABSurfer outperformed both of them with an average Dice Similarity Coefficient 0.034 higher than FastSurferVINN and 0.052 higher than FreeSurfer.

\begin{table}[htb]
\centering
\small
\begin{tabular}{llll}
\toprule
{} &      \textbf{TABSurfer} &     FastSurfer &     FreeSurfer \\
\midrule
DSC ↑ &  \textbf{0.792 ± 0.012} &  0.758 ± 0.014 &  0.740 ± 0.009 \\
ASSD ↓  &  \textbf{0.661 ± 0.129} &  0.724 ± 0.048 &  0.858 ± 0.102 \\
\bottomrule
\end{tabular}
\caption{Metrics from comparing TABSurfer, FastSurferVINN, and FreeSurfer to a manual reference.}
\label{table:manual}
\end{table}

\section{Discussion and Conclusion}
\label{sec:typestyle}

This study presents TABSurfer, a novel 3D patch-based CNN-Transformer hybrid deep learning model for the task of subcortical segmentation. TABSurfer demonstrates both qualitative and quantitative improvements over existing traditional and deep learning tools across multiple datasets with accelerated processing times. These results showcase the advantages of both our hybrid architecture and 3D patch-based approach.

When evaluated against the FreeSurfer ground truth, TABSurfer consistently achieved strong metrics, surpassing the benchmark, FastSurferVINN, which struggled to reach the same performance. Qualitatively, we also observed higher quality in TABSurfer's segmentations.

We then verified TABSurfer's accuracy on a manual reference, outperforming both FreeSurfer and FastSurferVINN considerably. Although overall performance was lower than on the FreeSurfer ground truths, this discrepancy can be attributed to the rougher contours in the manual ground truths. While expert human annotators can be more precise in certain areas, manual segmentations are noisier overall and less reproducible. TABSurfer generates smooth contours while attaining a stronger grasp of the anatomy over FreeSurfer and FastSurferVINN, as shown in our higher metrics.

We improve on the state-of-the-art deep learning methods in two areas. First, our 3D patch-based approach preserves more intricate spatial relationships within the continuity of the anatomy compared to a 2D slice approach like in FastSurferVINN. Our chosen patch size of 96 x 96 x 96 is large enough to retain adequate global context while remaining within reasonable computational constraints. Second, we improve on the standard CNN-based architectures with the addition of a Transformer, aiding in the further extraction of context and long-range dependencies despite the limited local view of each patch. By reducing the sizes of the memory-intensive convolutional layers while expanding the more computationally efficient Transformer module, we enable the model to process both a large patch size and a substantial number of classes on standard hardware. 

Our deep learning approach to image segmentation presents an advantage over traditional methods through the training data as well. By training with augmentation on ten diverse datasets with even age and gender distributions, TABSurfer can adapt to and smooth over greater noise in the inputs. On the other hand, traditional approaches like FreeSurfer can be more sensitive to such variations in quality. This enhanced reliability is particularly crucial for applications that rely on precise structural analyses of subcortical regions. 

For a more comprehensive assessment of TABSurfer, future studies should target generalizability and reliability by evaluating on additional datasets, more scans of varying resolutions and quality, and test-retest experiments. 

While this study provides promising results, there are still areas for improvement going forward. Due to the large dimensions of intermediate tensors and gradients, stored mostly by the convolutional layers, the current model is computationally expensive to train, requiring over 16 GiB of GPU memory when using a batch size of 2. TABSurfer is also slower than FastSurfer, primarily due to our model's increased depth. On a GPU, TABSurfer typically takes over 70 seconds to segment 32 classes, whereas FastSurfer can segment 95 regions in under a minute. By improving model efficiency, we can better explore the capabilities of this architecture on more classes to handle whole brain segmentation. Future works should experiment with different patch sizes and model dimensions to further enhance both utility and performance.


\section{Compliance with ethical standards}
\label{sec:ethics}
This research study was conducted retrospectively using human subject data made available in open access by the Australian Imaging Biomarkers and Lifestyle Study of Ageing (AIBL), Frontotemporal Lobar Degeneration Neuroimaging Initiative (NIFD), Information eXtraction from Images (IXI), Open Access Series of Imaging Studies-1 (OASIS-1), Open Access Series of Imaging Studies-2 (OASIS-2), Southwest University Adult life-span Dataset (SALD), Southwest University Longitudinal Imaging Multimodal Brain Data Repository (SLIM), Parkinson’s Progression Markers Initiative (PPMI), SchizConnect (SchizConnect), Consortium for Reliability and Reproducibility (CoRR), and MindBoggle-101 datasets. Ethical approval was not required as confirmed by the license attached with the open access data.

\section{Acknowledgments}
\label{sec:acknowledgments}

No funding was received for conducting this study and there are no relevant financial or non-financial interests to disclose. We acknowledge Dr. Tal Nuriel for proofreading this paper. 


\bibliographystyle{IEEEbib}
\bibliography{strings,refs}

\end{document}